\documentclass[lettersize,journal]{IEEEtran}
\usepackage{amsmath,amsfonts}
\usepackage{algorithm}
\usepackage{array}
\usepackage[caption=false,font=normalsize,labelfont=sf,textfont=sf]{subfig}
\usepackage{textcomp}
\usepackage{stfloats}
\usepackage{url}
\usepackage{verbatim}
\usepackage{graphicx}
\usepackage{cite}
\usepackage{booktabs}
\usepackage{stfloats}
\usepackage{multirow}
\usepackage{amsmath}
\usepackage{amssymb}
\usepackage{array}
\newcolumntype{C}[1]{>{\centering\let\newline\\\arraybackslash\hspace{0pt}}m{#1}}
\usepackage{url}
\usepackage{color,soul}
\usepackage{multirow}
\usepackage{hhline}
\usepackage{comment}
\usepackage{rotating}
\usepackage{optidef}
\usepackage{bbm}
\usepackage{enumitem}
\newcounter{descriptcount}

\usepackage{xcolor}
\usepackage{float}
\usepackage{caption}
\usepackage{subcaption}
\usepackage{stfloats}
\usepackage{url}
\usepackage{bbm}  
\usepackage{arydshln}
\usepackage{cite}
\usepackage{algpseudocode}
\usepackage[T1]{fontenc} 
\usepackage[utf8]{inputenc}
\usepackage{times}
\usepackage{orcidlink}
\usepackage{hyperref}
\hypersetup{hidelinks}
\usepackage{graphicx}
\usepackage{subcaption}

\usepackage[acronym]{glossaries}
\newacronym{cwnd}{W}{Congestion Window}
\newacronym{aimd}{AIMD}{Additive Increase Multiplicative Decrease}
\newacronym{tcp}{TCP}{Transmission Control Protocol}
\newacronym{rtt}{RTT}{Round Trip Time}
\newacronym{cca}{CCA}{Congestion Control Algorithm}
\newacronym{bbr}{BBR}{Bottleneck Bandwidth and Round-trip propagation time}
\newacronym{fcc}{FCC}{Federal Communications Commission}
\newacronym{bdp}{BDP}{Bandwidth-Delay Product}
\newacronym{dtc}{DTC}{Direct to Cell}
\newacronym{mbb}{MBB}{Make Before Break}
\newacronym{3gpp}{3GPP}{3rd Generation Partnership Project}
\newacronym{5g}{5G}{Fifth Generation}
\newacronym{6g}{6G}{6th-Generation}
\newacronym{ai}{AI}{Artificial Intelligence}
\newacronym{cnn}{CNN}{Convolutional Neural Network}
\newacronym{cogsat}{CogSat}{Cognitive Satellite}
\newacronym{cr}{CR}{Cognitive Radio}
\newacronym{crsn}{CRSN}{Cognitive Radio Sensor Networks}
\newacronym{csa}{CSA}{Concurrent Spectrum Access}
\newacronym{csi}{CSI}{Channel State Information}
\newacronym{dlr}{DLR}{Deep Learning Reinforcement}
\newacronym{dsa}{DSA}{Dynamic Spectrum Allocation}
\newacronym{dsrc}{DSRC}{Dedicated Short-Range Communications}
\newacronym{fl}{FL}{Federated Learning}
\newacronym{geo}{GEO}{Geostationary Equatorial Orbit}
\newacronym{iobt}{IoBT}{Internet of Battle Things}
\newacronym{iot}{IoT}{Internet of Things}
\newacronym{isl}{ISL}{Inter-Satellite Link}
\newacronym{itu}{ITU}{International Telecommunication Union}
\newacronym{leo}{LEO}{Low Earth Orbit}
\newacronym{lstm}{LSTM}{Long Short-Term Memory}
\newacronym{madrl}{MADRL}{Multi-Agent Deep Reinforcement Learning}
\newacronym{mdp}{MDP}{Markov Decision Process}
\newacronym{meo}{MEO}{Medium Earth Orbit}
\newacronym{ml}{ML}{Machine Learning}
\newacronym{nfv}{NFV}{Network Function Virtualization}
\newacronym{ntn}{NTN}{Non-Terrestrial Networks}
\newacronym{osa}{OSA}{Opportunistic Spectrum Access}
\newacronym{pu}{PU}{Primary User}
\newacronym{qos}{QoS}{Quality of Service}
\newacronym{rl}{RL}{Reinforcement Learning}
\newacronym{rsma}{RSMA}{Rate Splitting Multiple Access}
\newacronym{satcom}{SatCom}{Satellite Communication}
\newacronym{siot}{SIoT}{Satellite Internet of Things}
\newacronym{sl}{SL}{Supervised Learning}
\newacronym{su}{SU}{Secondary User}
\newacronym{uav}{UAV}{Unmanned Aerial Vehicle}
\newacronym{usl}{USL}{Unsupervised Learning}
\newacronym{vsat}{VSAT}{Very Small Aperture Terminal}
\newacronym{wsn}{WSN}{Wireless Sensor Networks}
\newacronym{gps}{GPS}{Global Positioning Systems}
\newacronym{dsm}{DSM}{Dynamic Spectrum Management}
\newacronym{stn}{STN}{Satellite Terrestrial Network}
\newacronym{eirp}{EIRP}{Effective Isotropic Radiated Power}
\newacronym{ss}{SS}{Sectrum Sensing}
\newacronym{kpi}{KPI}{Key Performance Indicator}
\newacronym{ca}{CA}{Channel Availability}
\newacronym{sinr}{SINR}{Signal to Noise plus Interference Ratio}
\newacronym{snr}{SNR}{Signal to Noise Ratio}
\newacronym{inr}{INR}{Interference to Noise Ratio}
\newacronym{ipc}{IPC}{Interference Power Constraint}
\newacronym{drl}{DRL}{Deep Reinforcement Learning}
\newacronym{ddpg}{DDPG}{Deep Deterministic Policy Gradient}
\newacronym{rem}{REM}{Radio Environment Map}
\newacronym{sdn}{SDN}{Software Defined Network}
\newacronym{rnn}{RNN}{Recurrent Neural Network}
\newacronym{dl}{DL}{Deep Learning}
\newacronym{svm}{SVM}{Support Vector Machine}
\newacronym{dqn}{DQN}{Deep Q-Network}
\newacronym{dca}{DCA}{Dynamic Channel Allocation}
\newacronym{wan}{WAN}{Wide Area Network}
\newacronym{gnss}{GNSS}{Global Navigation Satellite System}
\newacronym{ppo}{PPO}{Proximal Policy Optimization}
\newacronym{vits}{VITS}{Very High Throughput Satellites}
\newacronym{llm}{LLM}{Large Language Model}
\newacronym{sdr}{SDR}{Software Defined Radio}
\newacronym{css}{CSS}{Cooperative Spectrum Sensing}
\newacronym{ieee}{IEEE}{Institute of Electrical and Electronics Engineers}
\newacronym{rssi}{RSSI}{Received Signal Strength Indicator}
\newacronym{wran}{WRAN}{Wireless Regional Area Network}
\newacronym{cstn}{CSTN}{Cognitive Satellite Terrestrial Network}
\newacronym{bs}{BS}{Base Stations}
\newacronym{noma}{NOMA}{Non-Orthogonal Multiple Access}
\newacronym{cicstn}{CI-CSTN}{Cooperative Integrated-Cognitive Satellite Terrestrial Network}
\newacronym{hts}{HTS}{High Throughput Satellite}
\newacronym{mimo}{MIMO}{Multiple Input Multiple Output}
\newacronym{fss}{FSS}{Fixed Satellite Service}
\newacronym{mss}{MSS}{Mobile Satellite Service}
\newacronym{bss}{BSS}{Broadcasting Satellite Service}
\newacronym{mifr}{MIFR}{Master International Frequency Register}
\newacronym{wrc}{WRC}{World Radiocommunication Conference}
\newacronym{ngso}{NGSO}{Non-Geostationary Satellite Orbits}
\newacronym{mtc}{MTC}{Machine-Type Communications}
\newacronym{nbiot}{NB-IoT}{Narrowband Internet of Things}
\newacronym{etsi}{ETSI}{European Telecommunications Standards Institute}
\newacronym{bsm}{BSM}{Broadband Satellite Multimedia}
\newacronym{haps}{HAPS}{High-Altitude Platform Stations}
\newacronym{embb}{eMBB}{Enhanced Mobile Broadband}
\newacronym{mmtc}{mMTC}{Massive Machine Type Communications}
\newacronym{urllc}{URLLC}{Ultra-Reliable Low-Latency Communications}
\newacronym{ebu}{EBU}{European Broadcasting Union}
\newacronym{dth}{DTH}{Direct-to-Home}
\newacronym{acma}{ACMA}{Australian Communications and Media Authority}
\newacronym{cept}{CEPT}{European Conference of Postal and Telecommunications Administrations}
\newacronym{citel}{CITEL}{Inter-American Telecommunication Commission}
\newacronym{apt}{APT}{Asia-Pacific Telecommunity}
\newacronym{aws}{AWS}{Amazon Web Services}
\newacronym{esa}{ESA}{European Space Agency}
\newacronym{qpsk}{QPSK}{Quadrature Phase Shift Keying}
\newacronym{qam}{QAM}{Quadrature Amplitude Modulation}
\newacronym{tvws}{TVWS}{Television White Spaces}
\newacronym{darpa}{DARPA}{Defense Advanced Research Projects Agency}
\newacronym{iss}{ISS}{International Space Station}
\newacronym{scan}{SCaN}{Space Communications and Navigation}
\newacronym{knn}{KNN}{K-Nearest Neighbours}
\newacronym{pca}{PCA}{Principal Component Analysis}
\newacronym{gpt}{GPT}{Generative Pre-trained Transformer}
\newacronym{gan}{GAN}{Generative Adversarial Network}
\newacronym{fpga}{FPGA}{Field-Programmable Gate Array}
\newacronym{gpp}{GPP}{General-Purpose Processor}
\newacronym{vnf}{VNF}{Virtual Network Function}
\newacronym{mec}{MEC}{Multi Access Edge Computing}
\newacronym{usa}{USA}{United States of America}
\newacronym{uk}{UK}{United Kingdom}
\newacronym{vae}{VAE}{Variational Autoencoder}
\newacronym{los}{LoS}{Line of Sight}
\newacronym{bbm}{BBM}{Break Before Make}
\newacronym{acm}{ACM}{Adaptive Coding and Modulation}
\newacronym{ho}{HO}{Handover}
\newacronym{rb}{RB}{Resource Block}
\newacronym{prach}{PRACH}{Physical Random Access Channel}
\newacronym{isp}{ISP}{Internet Service Provider}
\newacronym{mcs}{MCS}{Modulation and Coding Scheme}
\newacronym{ack}{ACK}{Acknowledgement}
\newacronym{dnn}{DNN}{Deep Nural Network}
\newacronym{lnn}{LNN}{Linear Neural Network}
\newacronym{lora}{LoRA}{Low Rank Adaptation}
\newacronym{slm}{SLM}{Small Language Model}
\newacronym{vram}{VRAM}{Video Random Access Memory}
\newacronym{ffn}{FFN}{Feed-Forward Network}

\makeglossaries

\begin{document}

\title{Improving TCP BBR for LEO Satellites Using Lightweight Language Models}

\title{Adapting Small Language Models for Improved BBR Pacing in Low Earth Orbit Satellite Internet}
\title{Small Language Model-based Control for BBR over Low  Earth Orbit Satellite Internet}
\author{
    Rakshitha De Silva \orcidlink{0000-0002-7194-7619}, 
    Shiva Raj Pokhrel \orcidlink{0000-0001-5819-765X}~\textit{{Senior~Member,~IEEE}} and
    Jonathan Kua \orcidlink{0000-0001-9699-9418}~\textit{{Member,~IEEE}}
    \thanks{This work is supported by SmartSat CRC, whose activities are funded by the Australian Government’s CRC Program.\\
    R.~De~Silva, S.~R.~Pokhrel and J.~Kua are with the IoT \& Software Engineering Research Lab, School of Information Technology, Deakin University, Geelong, VIC 3125, Australia (e-mail: \href{mailto:rakshitha.desilva@deakin.edu.au}{rakshitha.desilva@deakin.edu.au}; \href{mailto:shiva.pokhrel@deakin.edu.au}{shiva.pokhrel@deakin.edu.au}; \href{mailto:jonathan.kua@deakin.edu.au}{jonathan.kua@deakin.edu.au}).\\
    }
}

\markboth{Journal of \LaTeX\ Class Files,~Vol.~X, No.~X, Month~Year}%
{Shell \MakeLowercase{\textit{et al.}}: A Sample Article Using IEEEtran.cls for IEEE Journals}


\maketitle

\vspace{-200mm}
\begin{abstract}


Low Earth Orbit (LEO) satellite Internet introduces rapid path variability, intermittent capacity shifts, and non-terrestrial delay dynamics that challenge transport-layer congestion control. Although Bottleneck Bandwidth and Round-trip propagation time (BBR) achieves high throughput in such environments, its aggressive bandwidth probing can cause excessive retransmissions and unstable pacing over LEO links. This paper presents a global experimental evaluation of BBR over a SpaceX Starlink testbed spanning six geographically distributed AWS endpoints and compares its behaviour against Cubic, Vegas, and Hybla under isolated and competing traffic scenarios. The measurements show that BBR consistently delivers superior throughput but incurs significantly higher retransmission overhead, revealing a critical throughput--stability trade-off in LEO satellite Internet. To address this limitation, we propose a Small Language Model (SLM)-guided BBR adaptation framework that learns phase-safe pacing-gain decisions from real Starlink traces. The framework combines a structured BBR state encoder, LoRA-based parameter-efficient fine-tuning, and a constrained networking head to generate feasible pacing actions with low inference latency. Evaluation using GPT-2, T5, GPT-Neo, and SmolLM2 shows that lightweight SLMs can retain BBR's throughput advantage while substantially reducing retransmissions, with performance comparable to larger language models but at much lower computational cost. 

\end{abstract}


\IEEEpeerreviewmaketitle

\section{Introduction}

\acrfull{leo} satellite mega-constellations~\cite{dou2025unleashing} have emerged as a cornerstone of next-generation global communications, enabling wide-area broadband with significantly lower latency than traditional satellite systems.
SpaceX Starlink represents the largest and most mature constellation to date, alongside efforts such as Eutelsat OneWeb and Amazon Kuiper. 
Google's \acrfull{bbr} \acrfull{cca}~\cite{bbrv3} marks a paradigm shift in \gls{tcp} congestion control by explicitly modeling bottleneck bandwidth and propagation delay rather than relying on loss-based detection.
Its latest iteration, BBRv3~\cite{bbrv3}\footnote{For the rest of the paper, if not stated otherwise, \gls{bbr}v3 is referred to as \gls{bbr}.}, refines probing and pacing to optimize throughput, latency, and fairness across diverse conditions. 
In parallel, \acrfullpl{slm}, a subclass of \acrfullpl{llm} designed to operate under strict compute, memory, and latency constraints~\cite{wang2025comprehensive}, retain meaningful language understanding within tight hardware budgets, making them well suited to edge, mobile, and embedded settings~\cite{nguyen-etal-2025-survey} and, in particular, to network-driven tasks that demand real-time inference close to the data source~\cite{lin2025pushing}. 

The recent surge in mega constellations has prompted industry and academia to optimize data transmission over satellite networks, generalizing it as a replacement and extension of well-established terrestrial networks~\cite{de2026understanding}.
Internet congestion control is a well-explored problem in the literature, yet inherited dynamics and transmission delays make it especially challenging in satellite networks~\cite{akyildiz2002tcp, taleb2006refwa}.
Many works have produced comparative analyses of \glspl{cca} over satellite networks~\cite{de2026understanding, tcpcomparision}, highlighting these associated challenges.
To this end, a significant number of contributions have been made in developing new \glspl{cca} for satellite Internet.
The most notable works include, but are not limited to: LeoCC~\cite{lai2025leocc}, TCP-Peach~\cite{akyildiz2002tcp}, REFWA~\cite{taleb2006refwa}, and StarQUIC~\cite{kamel2024starquic}.

Several studies have proposed refinements to \gls{bbr} to address fairness limitations and adapt its behavior to specific networking scenarios.
Modest BBR~\cite{zhang2018modest}, BBR-CWS~\cite{song2020bbr}, and BBR-ACD~\cite{mahmud2020bbr} mitigate \gls{bbr}'s aggressive probing by refining congestion window control and congestion detection, improving coexistence with loss-based flows and stability under shared bottlenecks.
More recent works, such as BBR-R~\cite{zheng2024bbr} and BBR-EFRA~\cite{njogu2023bbr}, focus on multi-flow competition and \gls{rtt} fairness, respectively, reducing bandwidth monopolization and unfairness across heterogeneous paths.
In parallel, delay-control adaptations similar to BBR-ES~\cite{han2026bbr} extension for low-latency applications. 

Recent efforts along this line leveraged \gls{drl} to replace or augment rule-based \glspl{cca}, as demonstrated in iCoCoA~\cite{donta2023icocoa}, enabling adaptive rate control under dynamic network conditions.
Several works have specifically targeted satellite networks, applying \gls{drl} to optimize congestion control in \gls{leo} constellations~\cite{zhu2025intelligent, yan2025handover}, \gls{sdn}-enabled satellite architectures~\cite{xing2022deep}, and deep-space communication scenarios~\cite{masood2022intelligent}, where high \glspl{rtt}, intermittent links, and handovers challenge conventional \glspl{cca}.
These approaches consistently show that learning-based controllers can better capture complex state--action relationships than handcrafted heuristics, leading to improved throughput, latency, and robustness.
More recently, the paradigm has expanded beyond \gls{drl} toward \glspl{llm}, with \gls{llm}-based queue management~\cite{pokhrel2026distilling}, cross-layer network tasks~\cite{wu2024netllm}, and \gls{llm}-driven cognitive spectrum adaptation~\cite{desilva2025adapt}, illustrating the potential of \glspl{llm} as high-level decision engines for network-aware control in next-generation satellite systems.
These advances, together with growing on-device and edge \gls{llm} deployments~\cite{lin2025pushing, chen2025towards}, motivate us to leverage \glspl{slm} to address elevated retransmission in \gls{bbr} over Starlink.
In this work, we have the following key contributions: 
\begin{itemize}
\item We design and implement a global Starlink performance testbed across six cities: Ohio, S ã o Paulo, London, Mumbai, Tokyo, and Sydney, on the \gls{aws} platform, and present a comprehensive empirical evaluation of \gls{bbr} under individual and competing uplink/downlink scenarios. 
\item We propose a \gls{slm}-based approach for smooth \gls{bbr} pacing gains, comprising a state encoder that projects \gls{bbr} numerical data into a language model-compatible feature space, a \gls{lora}-based \gls{rl} pipeline that distills language models for accelerated prediction of pacing gains, and a task-specific language model head that eliminates invalid action generation. 

\end{itemize}

We build a global, distributed testbed over the Starlink network, leveraging \gls{aws} infrastructure to study the empirical behavior of \gls{bbr}. 
The testbed expands to six \gls{aws} endpoints centered on Melbourne, Australia, and we compare \gls{bbr} performance with three leading \glspl{cca}: Cubic \cite{ha2008cubic}, Vegas \cite{brakmo2002tcp}, and Hybla \cite{caini2004tcp}. 
The observations highlight a throughput advantage, as \gls{bbr} significantly outperforms all three baseline \glspl{cca}  in both uplink and downlink, under isolated and competitive scenarios, fully utilizing the available Starlink Ku-band. 
The larger congestion window achieved through the probing mechanism is the primary facilitator of this advantage.  
Moreover, this throughput advantage is realized within comparatively similar \gls{rtt} and \gls{rtt}-variance margins to other \glspl{cca}. 
Compared to benchmark \glspl{cca}, the primary downside we observed is the notably high number of \gls{tcp} retransmissions, in all four test scenarios, across all six locations. 
This is a drawback resulting from the intentional aggressive probing for high bandwidth in the Starlink network, which often leads to inaccurate bottleneck estimates in dynamic network paths. 
\newline Exploiting the collected data from the distributed testbed, we proposed a \gls{slm} driven \gls{bbr} approach, reducing the inherent aggressiveness through smooth pacing gains. 
Due to the lightweight deployment advantages and inherited generalization capabilities of \glspl{slm}, our developed model represents a further step towards intelligence-driven network control for satellite Internet.

 Moreover, we formulate the BBR CCA problem as offline return-conditioned policy learning over measurement traces. For tractability, we evaluate the proposed approach with four celebrated \glspl{slm}, viz., 
GPT-2\footnotemark[1], T5\footnotemark[2], GPT-Neo\footnotemark[3], SmolLM2\footnotemark[4], and an \gls{llm} (LLaMA~3.2\footnotemark[5]  $>$3B parameters).  For reproducibility, we provide the complete implementation, including preprocessing, SLM fine-tuning, reward construction, pacing-gain inference and evaluation scripts at: \url{https://github.com/MPTCP-FreeBSD/lm-bbr-starlink.git}. 

\footnotetext[1]{\url{https://huggingface.co/openai-community/gpt2}}
\footnotetext[2]{\url{https://huggingface.co/docs/transformers/en/model_doc/t5}}
\footnotetext[3]{\url{https://www.eleuther.ai/artifacts/gpt-neo}}
\footnotetext[4]{\url{https://ollama.com/library/smollm2}}
\footnotetext[5]{\url{https://www.llama.com/}}

\section{Our Experiment: Observations and Findings}
\label{sec:testbed}

\subsection{Testbed Setup}
\begin{figure}[t]
\centerline{\includegraphics[width=0.9\columnwidth]{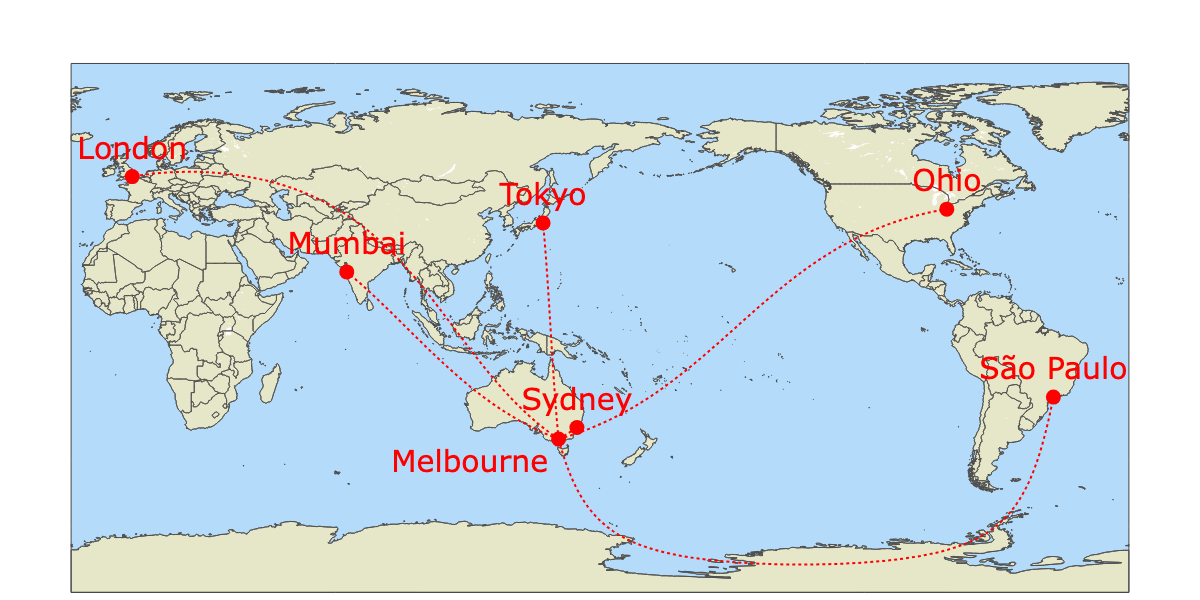}}
\vspace{-2mm}
\caption{Global testbed setup centered on Melbourne, expanding to AWS endpoints in six cities: Ohio, São Paulo, London, Mumbai, Tokyo, and Sydney. \label{fig:testbed}}
\vspace{-10pt} 
\end{figure}

As illustrated in Fig. \ref{fig:testbed}, we set up six servers in main cities distributed globally, namely Ohio, São Paulo, London, Mumbai, Tokyo, and Sydney. 
We leveraged \gls{aws} Linux cloud instances for this distributed server setup, and the local portion of the setup was based on the university premises in Melbourne, Australia. 
The user terminal consists of a Starlink latest-generation standard kit with a UTA-232 model dish. 
A Linux-based local server was connected to the Starlink terminal through a Category 5e Ethernet connection, creating a perpetual test environment. 
Furthermore, we assume the \gls{aws} cloud instances to provide a consistent network connection throughout the geographically distributed cloud instances. 
To test \gls{tcp} \gls{cca} performance over the testbed, we leverage \texttt{iperf3}, an open-source network testing tool used to measure the maximum achievable bandwidth and performance over network connections.
It provides detailed metrics on \gls{cca} performance such as throughput, retransmissions, congestion window, receiver advertised window, \gls{rtt}, and \gls{rtt}-variance, making it widely used for network diagnostics and benchmarking.

We implemented an automated measurement framework using \texttt{iperf3}, in which each remote server and local end was sequentially configured for specified \gls{cca}. 
We then performed forward and reverse tests for uplink and downlink for individual \gls{cca} data collections. 
To evaluate the \gls{cca} behavior as parallel and competing streams, we implemented isolated namespaces for each \gls{cca} in each Linux network instance and followed a similar data collection process. 
Special care was taken to synchronize server startup, ensuring data flushing before retrieval, and preserve isolation across \glspl{cca} to avoid inter-flow interference, thereby enabling fair and repeatable \gls{cca} benchmarking under identical network conditions.
Data was collected for a 300-second capture window for each \glspl{cca} in all four test scenarios, and the experiment was carried out during the first week of August 2025. 

\subsection{Evaluation: BBR over Starlink}



\begin{figure*}[!t]
\centering
\begin{minipage}{\textwidth}
    \centering
    \includegraphics[width=\linewidth]{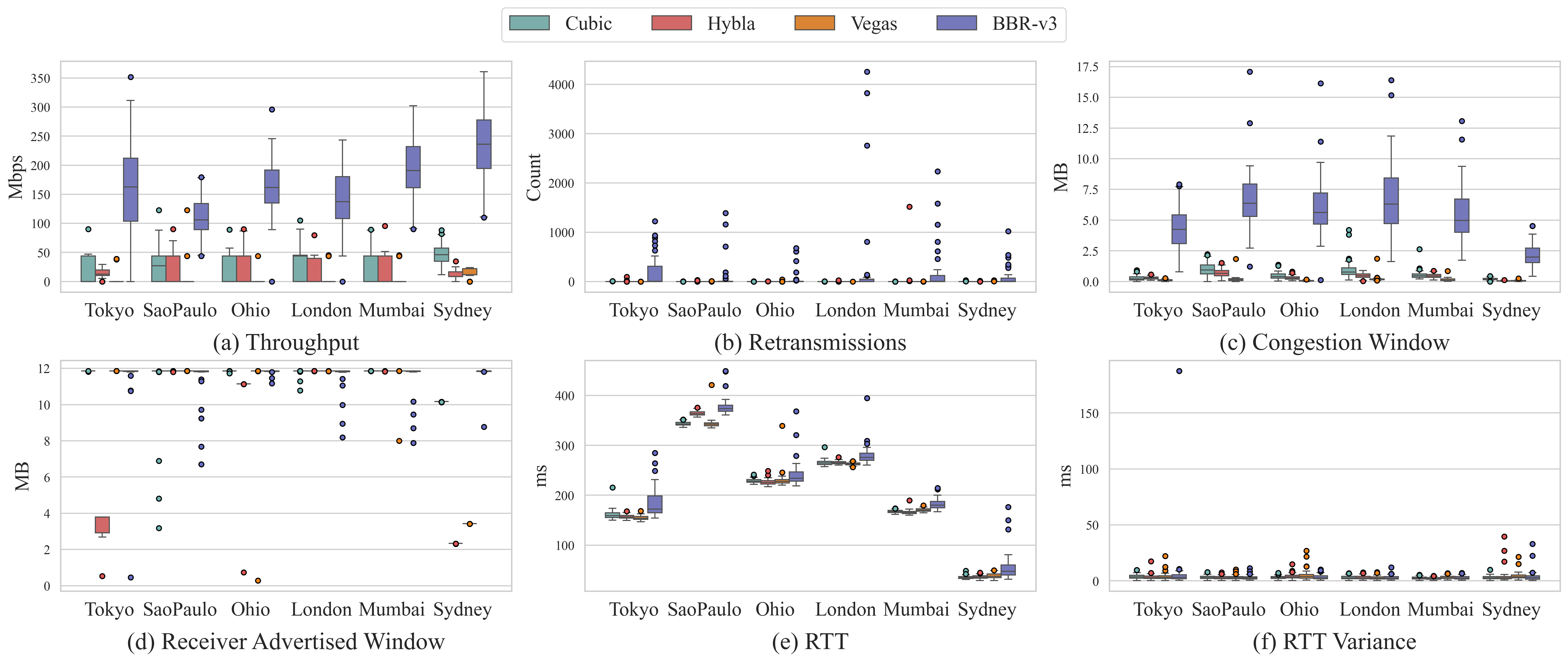}
    \captionof{figure}{Downlink data for independent CCA streams across the Starlink testbed.}
    \label{fig:ind_bbrdata_downlink}


    \includegraphics[width=\linewidth]{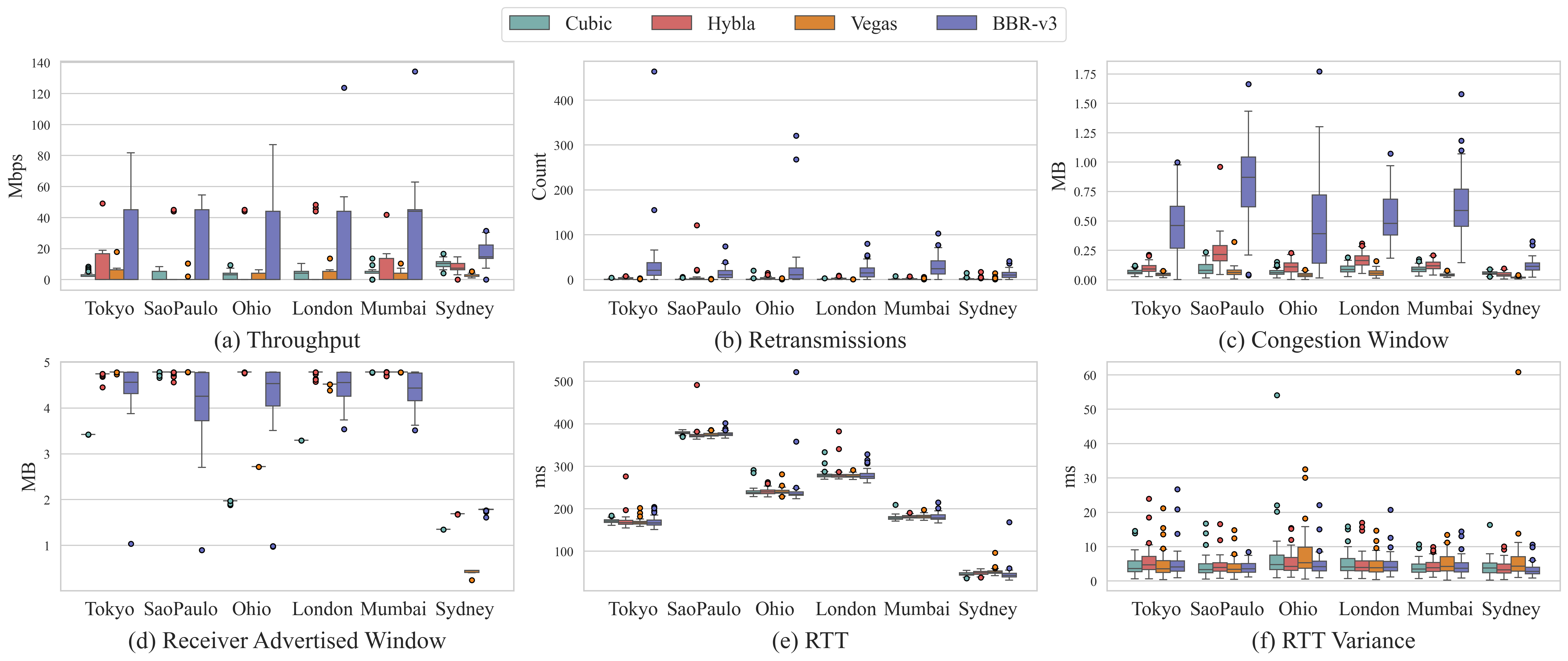}
    \captionof{figure}{Uplink data for independent CCA streams across the Starlink testbed.}
    \label{fig:ind_bbrdata_uplink}
\end{minipage}
\vspace{-10pt} 
\end{figure*}

\subsubsection{On Independent Noncompetitive Streams}
\label{subsection:testbed_data}
Across both independent download and upload streams, \gls{bbr} consistently achieved the highest throughput of all evaluated \glspl{cca} at every test location.
Downlink median ($Q_2$) values exceed $100$~Mbps in São~Paulo, and upper whiskers approach $350$~Mbps in Sydney (Fig.~\ref{fig:ind_bbrdata_downlink}(a)), while uplink throughput, though lower overall, still reaches upper whiskers exceeding $80$~Mbps (Fig.~\ref{fig:ind_bbrdata_uplink}(a)).
This performance came with markedly higher retransmissions than Cubic, Vegas, and Hybla, which all remained near-negligible.
In downlink, \gls{bbr}'s retransmissions peaked in Tokyo ($Q_3 > 300$) and Mumbai ($Q_3 > 100$), while in uplink they reached $Q_2 \in [20,30]$ with maxima near $70$, concentrated in high-\gls{rtt} locations (Fig.~\ref{fig:ind_bbrdata_downlink}(b), Fig.~\ref{fig:ind_bbrdata_uplink}(b)). 
\gls{bbr}'s congestion window was substantially larger than the other \glspl{cca} in both directions (min $Q_2 > 0.4$~MB in Sydney for uplink) with the highest overall variance, enabling higher \gls{bdp} utilization (Fig.~\ref{fig:ind_bbrdata_downlink}(c), Fig.~\ref{fig:ind_bbrdata_uplink}(c)). 
\glspl{rtt} were location-driven rather than \gls{cca}-driven, highest in São~Paulo ($Q_2 \approx 350$~ms) and lowest in Sydney ($Q_2 \approx 50$~ms), with \gls{rtt} variance remaining modest and similar across all \glspl{cca} (Fig.~\ref{fig:ind_bbrdata_downlink}(e),(f), Fig.~\ref{fig:ind_bbrdata_uplink}(e),(f)).

Despite its significantly higher throughput and retransmissions, \gls{bbr} maintained \gls{rtt} and \gls{rtt} variance comparable to the benchmark \glspl{cca}.
The steady, similar \gls{rtt} variance indicates that its retransmission overhead stemmed primarily from \gls{bbr}'s own aggressive probing rather than from Starlink's dynamics. This reflects a clear trade-off, \gls{bbr} operates closer to link capacity at the cost of stability, whereas the traditional loss and delay-based \glspl{cca} (Cubic, Vegas, Hybla) underutilize available capacity, confirming their limited effectiveness over Starlink. 
By relying on a model-based approach, \gls{bbr} bypasses the \gls{rtt}-dependence of conventional \glspl{cca}, making it an efficient choice for Starlink's dynamic \gls{leo} environment. Overall, \gls{bbr} demonstrates the strongest suitability for high-throughput operation over Starlink, though retransmission mitigation strategies are needed to improve its robustness.

\begin{figure*}[!t]
\centering
\begin{minipage}{\textwidth}
    \centering
    \includegraphics[width=\linewidth]{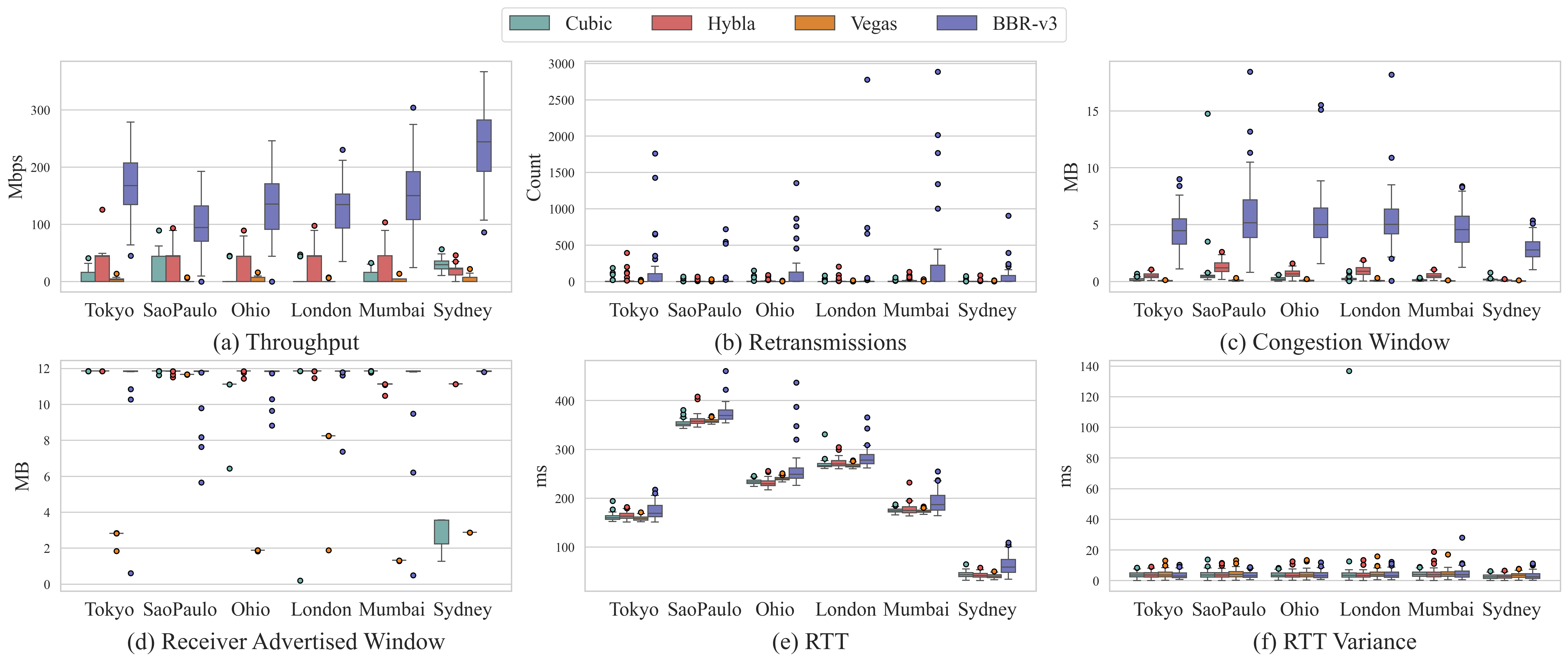}
    \captionof{figure}{Downlink data for parallel CCA streams over the global distributed Starlink testbed.}
    \label{fig:bbrdata_downlink}

    \includegraphics[width=\linewidth]{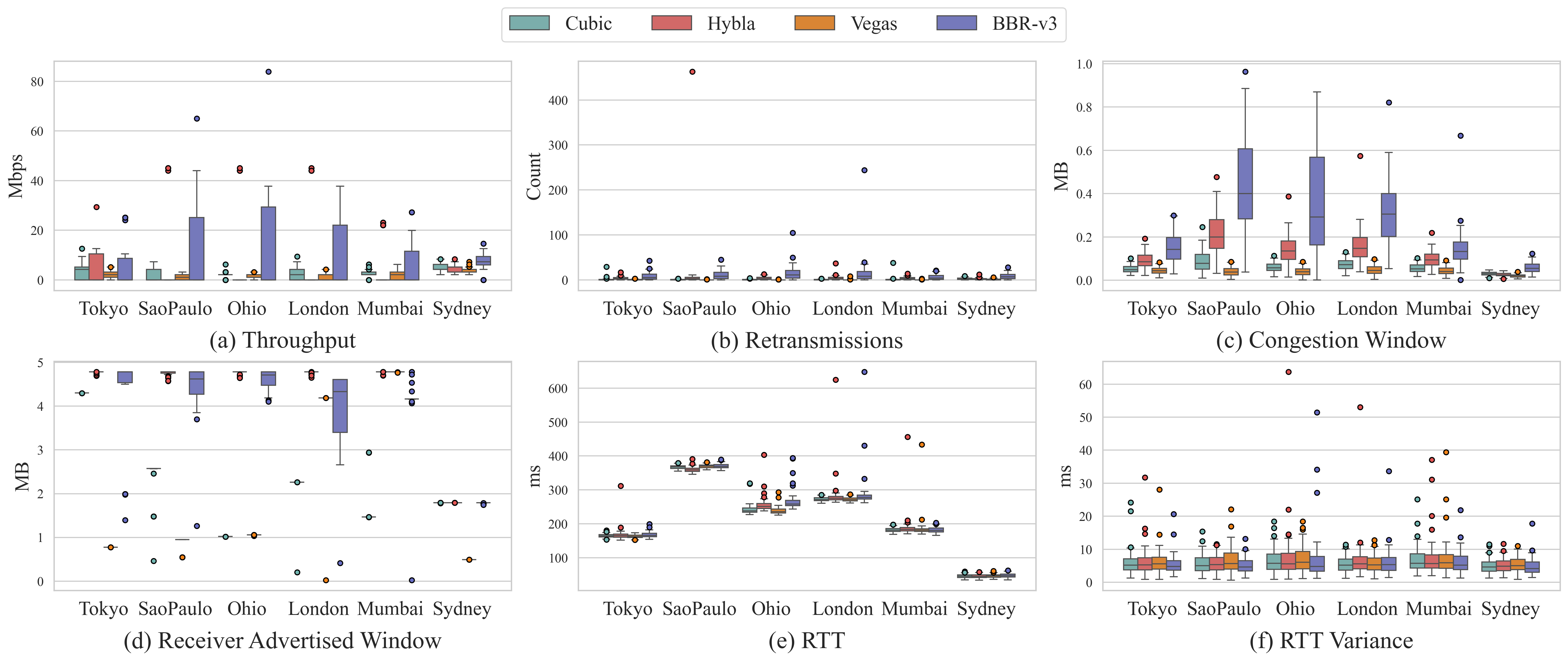}
    \captionof{figure}{Uplink data for parallel CCA streams over the global distributed Starlink testbed.}
    \label{fig:bbrdata_uplink}
\end{minipage}
\vspace{-14pt} 
\end{figure*}

\subsubsection{On Parallel Competitive Streams}
Under parallel, competitive stream conditions, \gls{bbr} consistently achieved the highest median throughput across all test sites in both directions, with downlink medians exceeding $130$~Mbps in London and $240$~Mbps in Sydney, with peaks above $350$~Mbps (Fig.~\ref{fig:bbrdata_downlink}(a)). 
In the uplink direction, upper whiskers exceeded $35$~Mbps, with Sydney achieving the best aggregated performance (Fig.~\ref{fig:bbrdata_uplink}(a)).
By contrast, the benchmark \glspl{cca} struggled to sustain capacity under contention.
Vegas frequently dropped to near-zero median downlink throughput, while Hybla outperformed Cubic and Vegas in selected locations ($Q_2 \approx 44$~Mbps downlink) yet degraded sharply in high-\gls{rtt} sites such as São~Paulo, London, and Ohio. 
This dominance came at the cost of the highest retransmissions in every location, with downlink upper whiskers exceeding $400$ packets in Mumbai and $200$ in Tokyo and Ohio, and uplink medians above $6$ packets reaching maxima beyond $30$ in high-\gls{rtt} locations (Fig.~\ref{fig:bbrdata_downlink}(b), Fig.~\ref{fig:bbrdata_uplink}(b)). 
\gls{bbr} also sustained the largest congestion windows (often $Q_2 > 5$~MB in downlink), enabling high in-flight data volumes while the other \glspl{cca} remained window-limited (Fig.~\ref{fig:bbrdata_downlink}(c), Fig.~\ref{fig:bbrdata_uplink}(c)). \gls{rtt} distributions showed only marginally higher variability for \gls{bbr} in downlink, with \gls{rtt} variance profiles following a similar pattern across all four \glspl{cca} and recording the highest \gls{rtt} in São~Paulo and the lowest in Sydney (Fig.~\ref{fig:bbrdata_downlink}(e),(f), Fig.~\ref{fig:bbrdata_uplink}(e),(f)).

\gls{bbr} consistently dominated competitive downlink and uplink capacity under parallel-stream conditions on Starlink, and similar to independent streams, its aggressive bandwidth probing delivered superior throughput at the cost of significantly increased retransmissions. The loss-based (Cubic) and delay-based (Vegas) \glspl{cca} instead sacrificed bandwidth efficiency to preserve stability and lower queuing pressure.
Hybla only partially mitigated \gls{rtt}-induced penalties and remained highly sensitive to contention, exhibiting unstable, burst-driven behavior. 
Notably, despite its marginally higher \gls{rtt} distribution in download streams, \gls{bbr} maintained \gls{rtt} and \gls{rtt} variance comparable to the benchmark \glspl{cca}. 
Overall, \gls{bbr} is best suited for capacity-dominant transfers over Starlink, whereas the more conservative \glspl{cca} favor predictable delay at the expense of throughput.

\vspace{-2mm}
\section{Proposed Finetuning SLM for BBR over Starlink}
\begin{figure*}[h]
\centerline{\includegraphics[width=1.7\columnwidth]{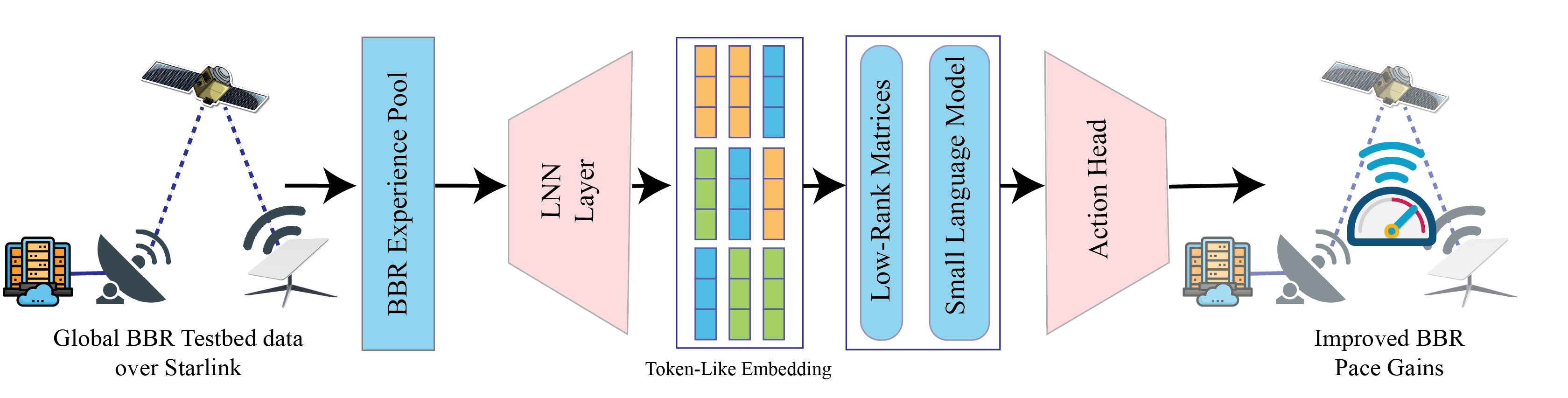}}
\vspace{-2mm}
\caption{An overview of the language model finetuning process for improved BBR. \label{fig:llm_setup}}
\vspace{-10pt} 
\end{figure*}

This section describes the data–processing pipeline developed to convert raw \texttt{iperf3} \gls{bbr} logs into structured data and \gls{slm} finetuning approach, as illustrated in Fig. \ref{fig:llm_setup}. 
The discussion is developed inline with \glspl{slm}, however, it is important to note that the proposed approach is compatible with \glspl{llm}, as emphasized in the result evaluation section later in this work. 

\begin{algorithm}[!t]
\caption{Creating Experience Pool from Starlink BBR Datasets}
\label{alg:processed_bbr_all_datasets}
\begin{algorithmic}[1]

\vspace{0.25em}
\State Concatenate \texttt{iperf3} BBR captures into a data-frame with $\{t,\, b_t,\, \tau_t,\, cwnd_t,\, rwnd_t,\, \mathrm{RTT}_t,\, \mathrm{RTTvar}_t\}$ values
\ForAll{dataset specifications $d \in \mathcal{D}_{BBR}$}

    \State Insert identifiers: Location flag -$L_j$, Stream flag -$S_k,$
    \State \textbf{Phase Detection:}
    \State Compute smoothed rolling mean throughput $\overline{b}(t)$.
    \State Compute deviation: Eq. (\ref{eq:std})
    \State Set thresholds $d_{\uparrow} \leftarrow \kappa \sigma_d$, $d_{\downarrow} \leftarrow -\kappa \sigma_d$
    \State Set initialize pace state \texttt{ProbeBW\_CRUISE} for all samples
    \State Scan for local maxima above $d_{\uparrow}$ to mark \texttt{ProbeBW\_UP}
    \State For each \texttt{ProbeBW\_UP} event, scan forward for local minima below $d_{\downarrow}$ to mark \texttt{ProbeBW\_DOWN}
    \State Mark subsequent 6 samples after \texttt{ProbeBW\_DOWN} as \texttt{ProbeBW\_CRUSE}

    \vspace{0.25em}
    \State \textbf{Phase-Constrained Gain Selection:}
    \State Compute per location/stream capacity estimate $B_{\max} \leftarrow \max_t bps(t)$
    \State Compute utilization proxy $\bar{B}_i\leftarrow b_i / \max_j b_j$

\ForAll{candidate gains $g\in\mathcal{A}_{\phi_i}$}
    \State \quad Estimate $\hat{B}_i(g)$ and $\hat{\tau}_i(g)$ trace-driven model
    \State \quad Compute $R(s_i,g)$
\EndFor
    \vspace{0.25em}
    \State Select $a_i^\star \leftarrow \arg\max_{g\in\mathcal{A}_{\phi_i}} R(s_i,g)$
    \State Append $(s_i,a_i^\star,R(s_i,a_i^\star))$ to $\mathcal{D}_{\mathrm{BBR}}$
\EndFor
\vspace{0.25em}
\State \Return $\mathcal{D}_{BBR} = \newline \{state \in \{L_j, S_k,t_i, b_i, \tau_i, cwnd_i, rwnd_i, \mathrm{RTT}_i, \mathrm{RTTvar}_i\}, \newline \mathcal{A} \in \{\mathcal{S}_{\mathrm{UP}}, \mathcal{S}_{\mathrm{DOWN}}, \mathcal{S}_{\mathrm{CRUISE}}\}, reward\}$

\end{algorithmic}
\end{algorithm}

\vspace{-2mm}
\subsection{Reinforcement Learning Pipeline}
As discussed in \S~\ref{sec:testbed}, \texttt{iperf3} \gls{bbr} logs contain per second measurements of throughput ($b$),
congestion window ($cwnd$), \gls{rtt} ($RTT$), retransmissions ($\tau$), receiver advertised window ($rwnd$), \gls{rtt} ($\mathrm{RTT}$) and \gls{rtt}-variance ($\mathrm{RTTvar}$). 
Thus, we consider a tuple of $\{t_i,\, b_i,\, \tau_i,\, cwnd_i,\, rwnd_i,\, \mathrm{RTT}_i,\, \mathrm{RTTvar}_i\}$ for each interval $i$, where $t_i$ is the respective time index. 
In \gls{bbr} bandwidth probing phases, pacing gain changes in three steps, in \texttt{ProbeBW\_DOWN}, it goes down to 0.9, in \texttt{ProbeBW\_CRUISE} and \texttt{ProbeBW\_REFILL}, it stabilizes at 1, and in \texttt{ProbeBW\_UP}, it goes up to 1.25 \cite{bbrv3}. 
This fixed pacing gain strategy may cause the \gls{cca}'s adjustment to lag when the bandwidth deviates, particularly in highly variable environments like \gls{leo} satellite networks \cite{yang2022bbrv2+, wang2024improved}.

\subsubsection{Discrete Pacing Gain Selection:}

Instead of the three-step pacing gain scheme, we introduce a granular discrete approach to counter the inherent aggressiveness. 
Let $\bar{b}_i$ denote a centered rolling–mean smoothed throughput estimate, and we define $d_i$ as:
\begin{equation}
d_i = b_i - \bar{b}_i ,
\label{eq:std}
\end{equation}
with standard deviation $\sigma_d = \mathrm{std}(d_i)$. 
An index sample is classified as \texttt{ProbeBW\_UP} if $d_i > 0.7\sigma_d$ and $d_i$ forms a local maximum, and
as \texttt{ProbeBW\_DOWN} if $d_i < -0.7\sigma_d$ and is a local minimum. 
A fixed pacing gain region of six sample cycles is inserted immediately following a \texttt{ProbeBW\_DOWN} event detection, aligning with the \gls{bbr} state machine. 

To model \gls{bbr} pacing behavior at a finer resolution, we discretize the gain around cruise pacing gain as $\mathcal{S}_{\mathrm{UP}} = \{1.05,\, 1.10, \, 1.15,\, 1.20,\, 1.25\}$, $\mathcal{S}_{\mathrm{DOWN}} = \{0.90,\, 0.92,\, 0.94,\, 0.96,\, 0.98\}$ and $\mathcal{S}_{\mathrm{CRUISE}} = 1.00 $. 
We leveraged a continuous gain estimate model derived as \cite{wang2024improved}:
\begin{equation}
\label{eq:continus_gain}
G^{\uparrow}_i = \frac{3}{\bar{B}_i+2}, \qquad
G^{\downarrow}_i = \frac{\bar{B}_i+1}{2},
\end{equation}
where $\bar{B}_i = b_i / \max_j b_j$ is a normalized rate–based throughput utilization estimate. 
Thus, we select gain as the nearest discrete value permitted by the detected macro–phase:
\begin{equation}
\label{eq:descrite_gain}
G_i =
\begin{cases}
\arg\min\limits_{g \in \mathcal{S}_{\mathrm{UP}}} |g - G^{\uparrow}_i|,
    & \mathrm{phase}_i = \mathrm{UP},\\[2pt]
\arg\min\limits_{g \in \mathcal{S}_{\mathrm{DOWN}}} |g - G^{\downarrow}_i|,
    & \mathrm{phase}_i = \mathrm{DOWN},\\[2pt]
1.00, & \mathrm{phase}_i = \mathrm{CRUISE}.
\end{cases}
\end{equation}

\subsubsection{Hybrid Utilization and Reward Construction:}
We define normalized throughput and retransmission reward components as:
\begin{align}
B_i &= \min\!\left(\frac{b_i}{B_i^{\mathrm{ref}}},\,1\right),\\
\bar{\tau}_i &= \tanh\!\left(\frac{\tau_i}{\bar{\tau}_i^{\mathrm{ref}}}\right).
\end{align}
where $B_i^{\mathrm{ref}}$ and $\bar{\tau}_i^{\mathrm{ref}}$ are interpreted as moving $95^{\mathrm{th}}$ percentile references for throughput and retransmissions where:
\begin{align}
B_i^{\mathrm{ref}} &= Q_{0.95}(b_{i-w:i+w}),\\
\bar{\tau}_i^{\mathrm{ref}} &= Q_{0.95}(\tau_{i-w:i+w}) + 1,
\end{align}
and $Q_p(\cdot)$ denotes the empirical percentile and $w$ is the window size. 
A saturating function is leveraged in encapsulating retransmission to eliminate the effect of a large spike in the captured \gls{bbr} data. 

\begin{table}[t]
\centering
\caption{Hyperparameter settings used in reward and loss construction.}
\label{tab:hyperparameters}
\begin{tabular}{ccl}
\hline
Parameter & Value & Description \\
\hline
$w$ & $10$ & Half-window for rolling percentile estimates \\
$\kappa$ & $0.7$ & ProbeBW phase-detection threshold scale \\
$\lambda_1$ & $0.5$ & Retransmission penalty weight \\
$\lambda_2$ & $0.1$ & Aggressive probe-up penalty weight \\
$\alpha$ & $1.5$ & Superlinear probe-loss growth factor \\
$\beta$ & $5$ & Softplus probe-activation sharpness \\
$\epsilon$ & $10^{-3}$ & Minimum non-zero loss floor \\
$\kappa_{\mathrm{down}}$ & $0.5$ & Probe-down retransmission reduction factor \\
\hline
\end{tabular}
\end{table}
Throughput may saturate with excessive queue buildups, and delay alone may remain low during short probing phases even when bandwidth is underutilized.  
Therefore, to represent congestion effects in the reward function, a hybrid utilization estimate embodying rate and delay is incorporated.
Let $\mathrm{RTT}_{\min} = \min_j \mathrm{RTT}_j$ considering a rolling window, and define the queuing delay
$qd_i = \max(\mathrm{RTT}_i - \mathrm{RTT}_{\min},\,0)$. 
A bounded queue reference can be computed as:
\begin{equation}
q_i^{\mathrm{ref}} = 
\min\!\left(Q_{0.95}(qd_{i-w:i+w}),\; 0.15\,\mathrm{RTT}_{\min}\right).
\end{equation}
Thus, the rate–based and delay–based utilizations are defined as:
\begin{align}
U_i^{\mathrm{rate}} &= B_i,\\
U_i^{\mathrm{delay}} &= \min\!\left(\frac{qd_i}{q_i^{\mathrm{ref}}},\, 1\right),
\end{align}
Thus, the hybrid utilization factor can be given as:
\begin{equation}
U_i = \max(U_i^{\mathrm{rate}},\, U_i^{\mathrm{delay}}).
\end{equation}

A penalty is introduced to penalize aggressive pacing when $G_i > 1.0$:
\begin{equation}
l_i = \max(G_i - 1,\, 0).
\end{equation}
Thus, the reward function is defined as:
\begin{equation}
\label{eq:reward}
r_i = B_i - \lambda_1 C_i - \lambda_2\, U_i\, l_i,
\end{equation}
where $\lambda_1 = 0.5$ and $\lambda_2 = 0.1$.
Thus, we define an experience pool dataset along the \gls{rl} framework as $\mathcal{D}_{BBR} = \{state, action, reward\}$, incorporating the states: $s_i \in \{L_{i,j}, S_{i,k},t_i, b_i, \tau_i, cwnd_i, rwnd_i, \mathrm{RTT}_i, \mathrm{RTTvar}_i\}$ where $L_{i,j}$ is a location flag and $S_{i,k}$ is a stream identification flag for the respective dataset of the four types collected with the distributed testbed. 
The action set is defined as:  $a_i \in \{\mathcal{S}_{\mathrm{UP}}, \mathcal{S}_{\mathrm{DOWN}}, \mathcal{S}_{\mathrm{CRUISE}}\}$.
The framework used in this process is further detailed in Algorithm \ref{alg:processed_bbr_all_datasets} and hyperparameters are defined in Table~\ref{tab:hyperparameters}. 

We select the pacing gain from the finite action space $ \mathcal{A}=$
\begin{equation}\{0.90,0.92,0.94,0.96,0.98,1.00,1.05,1.10,1.15,1.20,1.25\}. \end{equation}
To preserve BBR phase consistency, the feasible set is
\begin{equation} \mathcal{A}_{\phi_i}= \begin{cases} \{1.05,1.10,1.15,1.20,1.25\}, & \phi_i=\text{BW\_UP},\\ \{0.90,0.92,0.94,0.96,0.98\}, & \phi_i=\text{BW\_DOWN},\\ \{1.00\}, & \phi_i=\text{BW\_CRUISE}. \end{cases} \end{equation}
For each candidate $g\in\mathcal{A}_{\phi_i}$, the trace-driven models estimate the delivered throughput $\hat{B}_i(g)$ and retransmission cost $\hat{\tau}_i(g)$. The reward is 
\begin{equation} R(s_i,g)= \frac{\hat{B}_i(g)}{B_i^{\mathrm{ref}}} -\lambda_1\frac{\hat{\tau}_i(g)}{\tau_i^{\mathrm{ref}}} -\lambda_2 U_i\max(g-1,0), \end{equation}
where $U_i$ denotes hybrid utilization. The expert label is the feasible gain that maximizes this reward: 
\begin{equation} a_i^\star=\arg\max_{g\in\mathcal{A}_{\phi_i}} R(s_i,g). \end{equation}
Thus, the experience pool is \begin{equation} \mathcal{D}_{\mathrm{BBR}}= \{(s_i,a_i^\star,R(s_i,a_i^\star))\}_{i=1}^{N}, \end{equation} providing phase-safe expert labels for offline return-conditioned policy learning.

\begin{algorithm}[ht]
\caption{SLM-based BBR for Starlink Internet}
\label{algo:cogsat-LLM}
\begin{algorithmic}[1]
\Require Transformed structured $\mathcal{D}_{BBR}$ to trainable data, action sequence $\mathcal{A}$ and feature dimensions $n_{\text{f}}$, where $\mathbf{x_t} = \large[{R}_t, {s}_{t}^{(1)}, \ldots, {s}_{t}^{(n_{\text{s}})}, {a}_t)\large]^{\text{T}} \in \mathbb{R}^{n_{\text{f}}}$

\Statex \textbf{Part I - State Encoder:}
\State Initialize State tensor $\mathbf{S} \in \mathbb{R}^{B \times T \times n_{\text{f}} \times 1}$ with 
embedding dimension $d'$, batch size $B$ and sequence length $T$
\State Initialize FC layers 
$\{\mathrm{FC}_i(\cdot)\}_{i=1}^{n_{\text{f}}}$ with LeakyReLU activation
\State Reshape state tensor: $\mathbf{S} \leftarrow \mathrm{reshape}(\mathbf{S}, (B \cdot T, n_{\text{f}}, 1))$
\For{$i = 1$ to $n_{\text{f}}$}
    \State Extract scalar feature: $\mathbf{s}_i \leftarrow \mathbf{S}(i)$
    \State Encode feature using FC layer: $\mathbf{f}_i \leftarrow \mathrm{FC}_i(\mathbf{s}_i)$
    \State Reshape encoded feature: $\mathbf{F}_i \leftarrow \mathrm{reshape}(\mathbf{f}_i, (B, T, d))$
\EndFor
\State \Return 
$\{\mathbf{f}_1, \mathbf{f}_2, \ldots, \mathbf{f}_{n_{\text{f}}}\}$

\Statex \textbf{Part II - Low-Rank Adaptation:}
\State Initialize matrices $A$ and $B$ with random weights $\theta_{lr}$
\For {epoch = 1 : epochs}
    \For{data batch = 1 to data batches in $\mathcal{D}_{BBR}$}
        \State Get the forward pass model output
        \State Compute the loss Eq. (\ref{eq:loss})
        \State Update $\theta_{lr}$
    \EndFor
\EndFor
\State \Return $W_{\tau} \gets W_\theta + A_{\tau}B_{\tau}$

\Statex \textbf{Part III - Offline RL Policy with Language Model Head:}

\State Embed actions ($\mathbf{E}_a$), returns ($\mathbf{E}_r$), and time steps ($\mathbf{E}_t$) as trainable linear mapping

\State Add temporal embeddings:
$\mathbf{E}_a \leftarrow \mathbf{E}_a + \mathbf{E}_t,\;
 \mathbf{E}_r \leftarrow \mathbf{E}_r + \mathbf{E}_t$

\State Encode state sequence using \textbf{State Encoder}

\For{$i=1$ to $n_{\text{f}}$}
    \State $\mathbf{E}_{s_i} \leftarrow \mathrm{Embed}_{s_i}(\mathbf{F}_i) + \mathbf{E}_t$
\EndFor

\State Construct autoregressive token sequence in $\mathbf{x_t}^{\text{T}}$ format

\State Stack tokens over time to form input sequence $\mathbf{X}$
\State Truncate $\mathbf{X}$ to model context length and apply layer normalization

\State Create attention mask $\mathbf{M}$
\State Compute hidden states using PLM:

$\mathbf{H} \leftarrow \mathrm{PLM}(\mathbf{X}, \mathbf{M})$

\If{residual connection enabled}
    \State $\mathbf{H} \leftarrow \mathbf{H} + \mathbf{X}$
\EndIf

\State Extract logits corresponding to action positions
\State Predict action logits:
$\hat{\mathbf{A}} \leftarrow \mathrm{ActionHead}(\mathbf{H})$

\State \Return $\hat{\mathbf{A}}$
\end{algorithmic}
\end{algorithm}

\vspace{-2mm}
\subsection{State Encoder}
To transform \texttt{iperf3} generated \gls{bbr} state information into representations compatible with \gls{slm} architectures, we employ a structured feature encoding mechanism tailored to the heterogeneous requirements of the selected models. 
A series of fully connected layers is first used to process the scalar inputs produced by our Starlink-\gls{bbr} global testbed. 
\glspl{slm} impose varying context window constraints; for example, GPT-2 supports sequences up to 1024 tokens, and T5 operates with a 512-token span. 
Thus, dimensional alignment between the encoded features and the target model's token space is essential. 
Therefore, we incorporate a \gls{lnn} layer that systematically maps the extracted features into embedding dimensions compatible with each model’s context window. 
This design ensures architectural coherence and facilitates seamless integration with \gls{slm} backbones of different scales and configurations, converting the extracted features into a matching context window. 
Layer normalization is subsequently applied to the projected embeddings to enhance numerical stability, mitigating Covariate shift and promoting efficient optimization. 
Therefore, the state encoder approach prepares \gls{bbr} data over Starlink in a form optimally structured for transformer-based processing while maintaining computational efficiency and robustness across diverse language model architectures.

\vspace{-2mm}
\subsection{LoRA Adaptation}

To improve training efficiency, we adopt a \gls{lora} strategy within the Parameter-Efficient Fine-Tuning (PEFT) framework \cite{xu2023parameter}. 
This approach enables the reduction of parameters updated during training by freezing the majority of the pre-trained model weights of a language model. 
Let $\theta_{\text{tot}}$ denote the full parameter set of the \gls{slm}, which can be decomposed into frozen parameters $\theta_{F}$ and trainable parameters $\theta_{T}$, such that $\theta_{\text{tot}} = \theta_{F} + \theta_{T}$. 
The corresponding weight matrices for the respective parameters can be denoted as $W_{\text{tot}}$, $W_{F}$, and $W_{T}$.
Under the \gls{lora} approach, the trainable weight update $W_{T} \in \mathbb{R}^{p \times q}$ is approximated using a low-rank decomposition. 
Specifically, we assume a rank $r$ satisfying $r \ll \min\{p, q\}$ and express $W_{T}$ as the product of two matrices, $A \in \mathbb{R}^{p \times r}$ and $B \in \mathbb{R}^{r \times q}$, such that $W_{T} = AB$. 
During fine-tuning, only the parameters in $A$ and $B$ are updated, while a significantly large portion of the original model weights remain fixed. 
This notably reduces computational and memory overhead while preserving the representational capacity of the pretrained \glspl{slm} for more generalized decision-making. 

\vspace{-2mm}
\subsection{Language Model Head and Training}
State encoder representations are subsequently forwarded to a dedicated \emph{networking head}, which directly generates task-specific outputs.
However, \gls{llm} hallucination, a well-known and explored issue within the field \cite{huang2025survey}, can cause the generation of actions that fall outside the valid range.
As a solution to this, the networking head is implemented as a trainable linear projection layer that maps the internal feature representations of the \gls{slm}.
This matches the output to be within the desired space, in this work,  modified \gls{bbr} pacing gains. 
Moreover, since the model produces a single valid output within one inference step, the overall decision latency is significantly reduced.

Recall the \gls{rl}-based pipeline leading to the experience pool, where $D_{BBR} = \{ r_{t}, \mathbf{s_{t}}, \mathbf{a_{t}}\}_{t=1}^{L} \in \mathcal{D}_{BBR}$ and $D_{BBR}$ denote a trajectory at $L$, the episode length.
For each $D_{BBR}$ we replace $r_t \leftarrow R_{t} = \Sigma _{i=t}^{L}r_i$, considering the cumulative rewards at $s_t$. 
Furthermore, state components, and action-related information are discretized as $s_t = \{ s_{t}^{1},\dots, s_{t}^{g} \}$ and $a_t = \{ a_{t}^{1},\dots, a_{t}^{h} \}$ exposing the incorporated information. 
The \gls{slm} is fine-tuned to learn return distributions using this data representation, with randomly sampled data sequences as:
\begin{equation}
    \bar{D}_{BBR} = \{ R_{i}, s_{t}^{1},\dots , s_{t}^{g}, a_{t}^{1},\dots , a_{t}^{h}\}_{i=t-w+1}^{t} \in \mathcal{D}_{BBR}
\end{equation}
The optimization objective for training \gls{slm} action generation is defined in Eq.~(\ref{eq:loss}), where $L_{H}(\cdot)$ denotes the cross-entropy loss computed between the ground truth action $a_{t'}^{j}$ and the predicted action $\hat{a}_{t'}^{j}$. 
The loss is averaged across a prediction horizon of length $w$ and across all $m$ action dimensions.

\begin{equation}
L = \frac{1}{w} \sum_{t'=1}^{w} \sum_{j=1}^{m} L_{H}\big(a_{t'}^{j}, \hat{a}_{t'}^{j}\big)
\label{eq:loss}
\end{equation}

The objective of this training strategy is to enable the \gls{slm} to internalize the relationship between system states and their associated return distributions.
After training, the model is capable of synthesizing action sequences that are consistent with a desired performance objective. 
The overall process adapting \glspl{slm} to improve \gls{bbr} pacing gains is further detailed in Algorithm \ref{algo:cogsat-LLM}.

\vspace{-2mm}
\section{Result Evaluation}
\label{sec:result}












\begin{table}[!t]
\centering
\caption{Comparison of GPT-2, T5, GPT-Neo, and SmolLM2 SLMs}
\label{tab:model_comparison}
\renewcommand{\arraystretch}{1.25}
\setlength{\tabcolsep}{4pt}
\small
\begin{tabular}{@{}lcccc@{}}
\toprule
\textbf{Property} &
\textbf{GPT-2} &
\textbf{T5} &
\textbf{\shortstack{GPTNeo}} &
\textbf{\shortstack{SmolLM2}} \\
\midrule
Parameters        & 137\,M   & 237\,M   & 132\,M   & 388\,M \\
Vocabulary Size  & 50{,}257 & 32{,}128 & 50{,}304 & 49{,}152 \\
Hidden Size   & 768      & 768      & 768      & 960 \\
Layers        & 12 d.    & \shortstack{12 e.\,+\\12 d.} & 12 d. & 32 d. \\
Attention Heads  & 12       & 12       & 12       & 15 \\
FFN Size      & 3072     & 3072     & 3072     & 2560 \\
Max Seq.\ Len & 1024     & 512      & 2048     & 8192 \\
Activation    & GELU     & ReLU     & GELU     & SiLU \\
Positional Encoding    & Learned  & \shortstack{Rel.\\Buckets} & \shortstack{RoPE\\(25\%)} & RoPE \\
\bottomrule
\end{tabular}
    \vspace{-14pt} 
\end{table}

\subsection{Training and Validation}
We separate the collected data detailed in \S~\ref{sec:testbed} as training and testing, grouping Ohio, São Paulo, London, Mumbai, and Sydney for training. 
Leveraging this real-world \gls{bbr} data set, we evaluate the performance of our proposed approach using four \glspl{slm}: GPT-2, T5, GPT-Neo, and SmolLM2. 
The primary selection criterion for these \glspl{slm} was the number of parameters; accordingly, we selected models with fewer than 400~M trainable parameters for this study.
A summary comparison of the \glspl{slm} of interest is given in Table \ref{tab:model_comparison}. 
For the training and evaluation process, we used a workstation equipped with an Intel Xeon Gold 6346 CPU and two NVIDIA RTX 6000 GPUs, each with 96~GB of VRAM.
To ensure a stable and efficient training process, we adopt mini-batches of 20 samples combined with gradient accumulation. 
Gradient clipping is also applied to constrain magnitude growth during backpropagation, thus preventing gradient explosion.

\begin{figure}[!t]
\centering
\begin{minipage}{0.49\columnwidth}
    \centering
    \includegraphics[width=\linewidth]{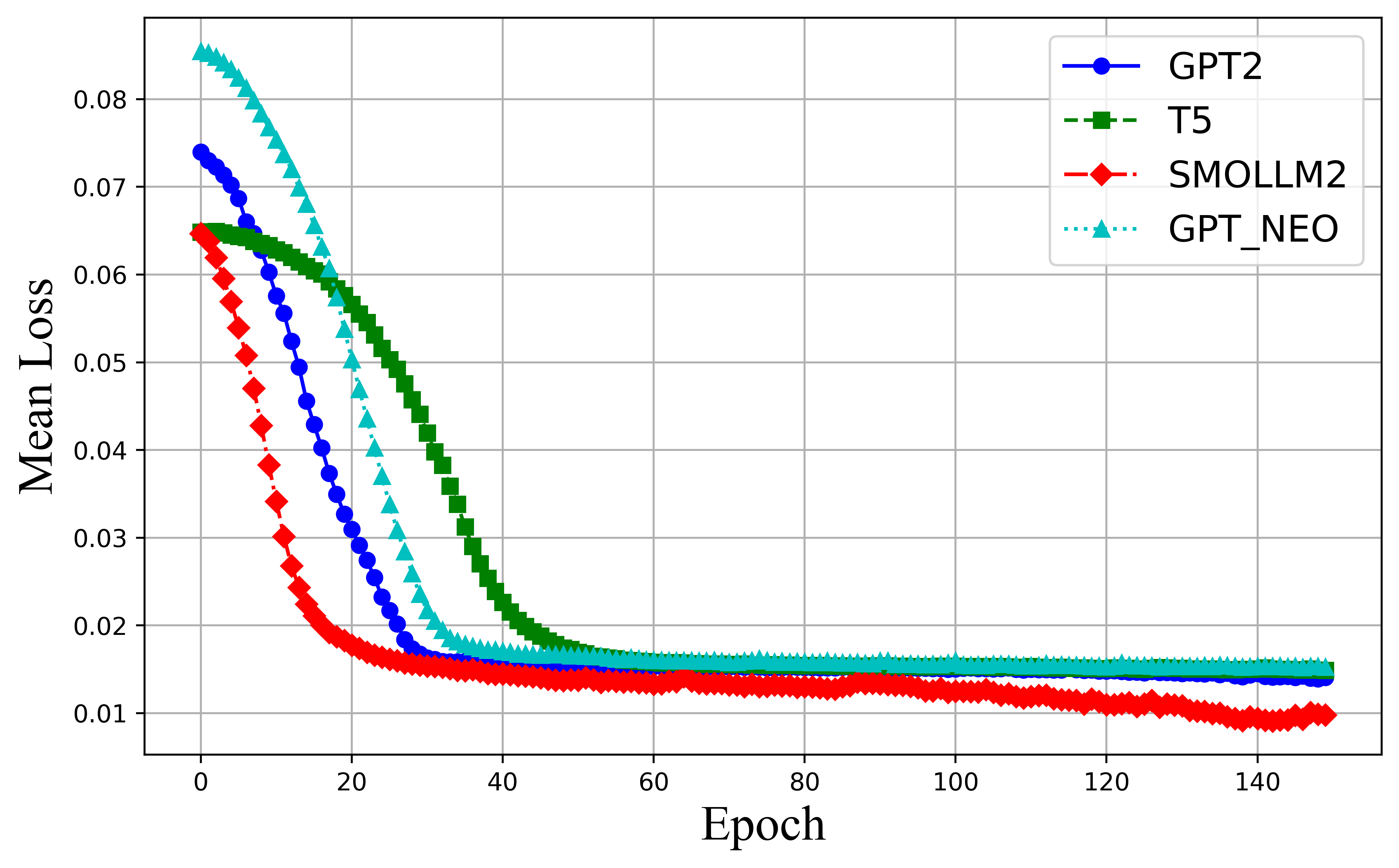}\\
    \small (a) Mean loss
\end{minipage}
\hfill
\begin{minipage}{0.49\columnwidth}
    \centering
    \includegraphics[width=\linewidth]{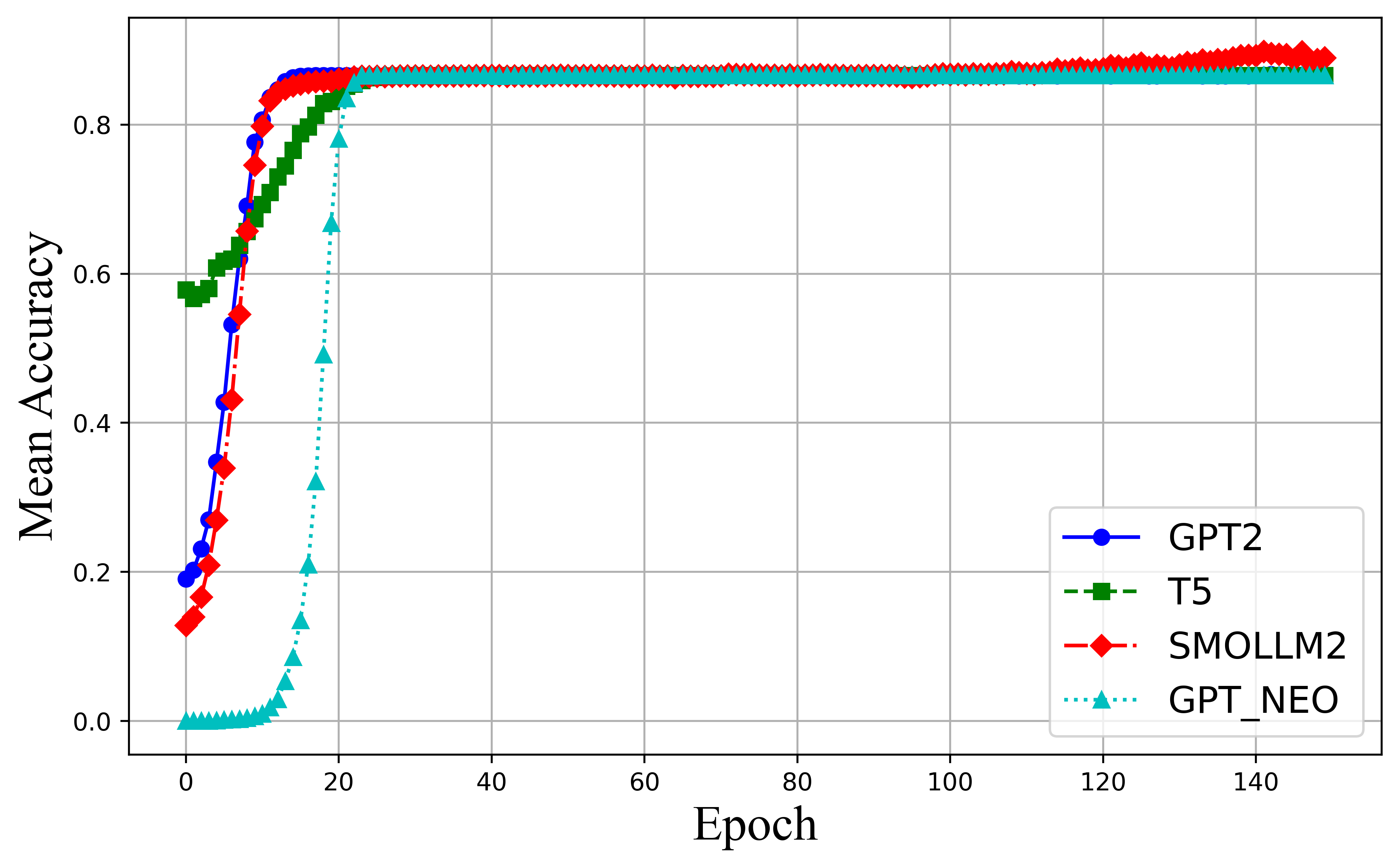}\\
    \small (b) Mean accuracy
\end{minipage}
\caption{Comparison of SLM training: (a) mean loss and (b) mean accuracy.}
\label{fig:training}
\vspace{-14pt} 
\end{figure}

\glspl{slm}' training performance was evaluated using cross-entropy loss for 150 epochs, as depicted in Fig. \ref{fig:training}(a). 
T5 was the slowest to reduce the mean loss, requiring more than 50 epochs to reach stability.
It was succeeded by GPT-Neo and GPT-2, and all three models exhibited similar loss patterns after convergence.
However, SmolLM2 deviated from the established pattern, reducing its mean loss relatively quickly and exhibiting fluctuations even after 100 epochs. 
Fig. \ref{fig:training}(b) shows the mean training accuracy of the \glspl{slm}. 
After 30 epochs, all four \glspl{slm} converged in terms of mean accuracy even though SmolLM2 displayed slight alterations, reflecting the loss pattern. 
Interestingly, GPT-2 and SmolLM2 recorded similar accuracy patterns in the beginning, increasing rapidly, followed by T5. 
Comparatively, GPT-Neo was the slowest model to converge in respect of mean accuracy. 




\begin{figure}[!t]
\centering
\begin{minipage}{0.49\columnwidth}
    \centering
    \includegraphics[width=\linewidth]{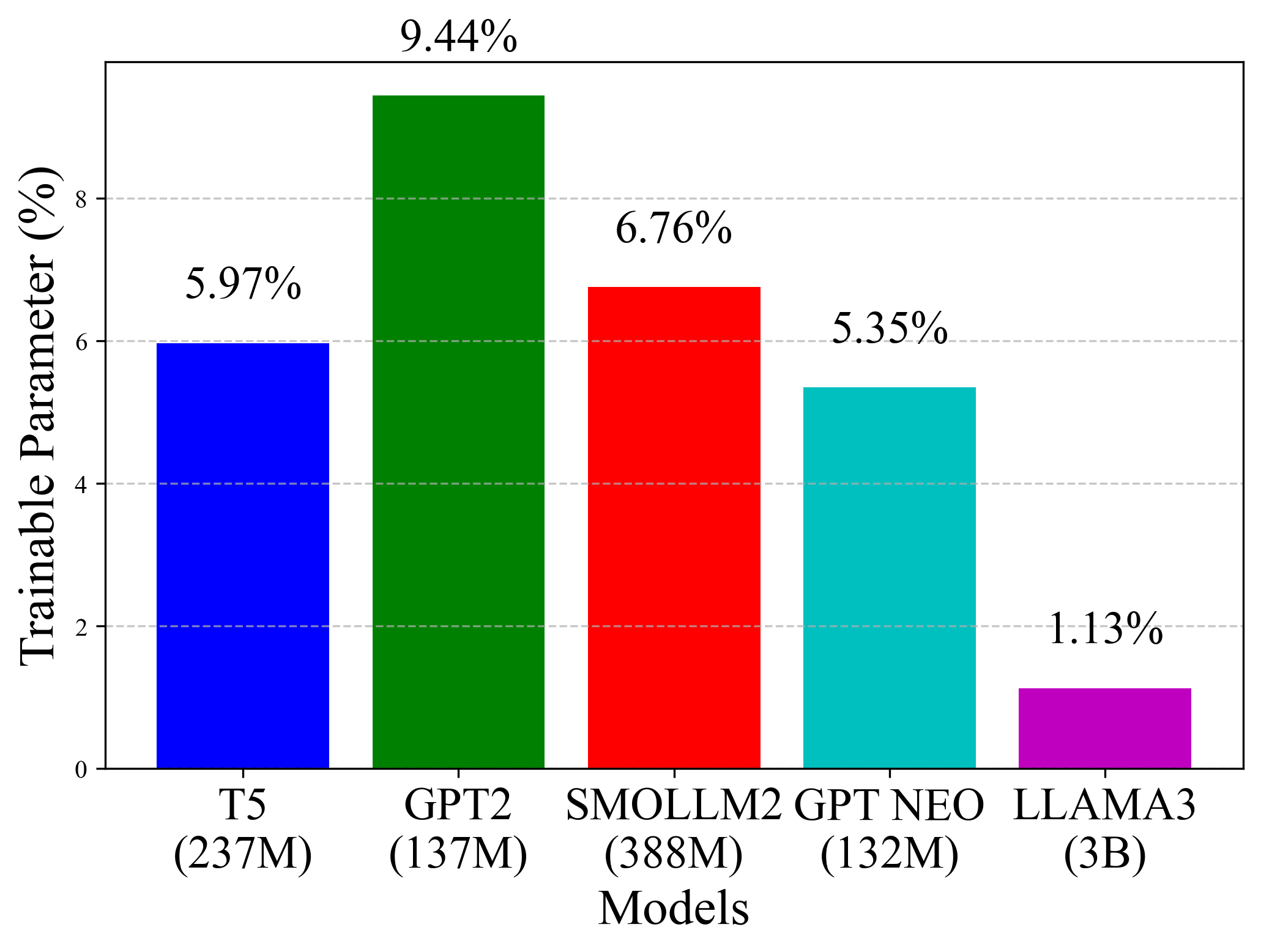}\\
    \small (a) Trainable parameters
\end{minipage}
\hfill
\begin{minipage}{0.49\columnwidth}
    \centering
    \includegraphics[width=\linewidth]{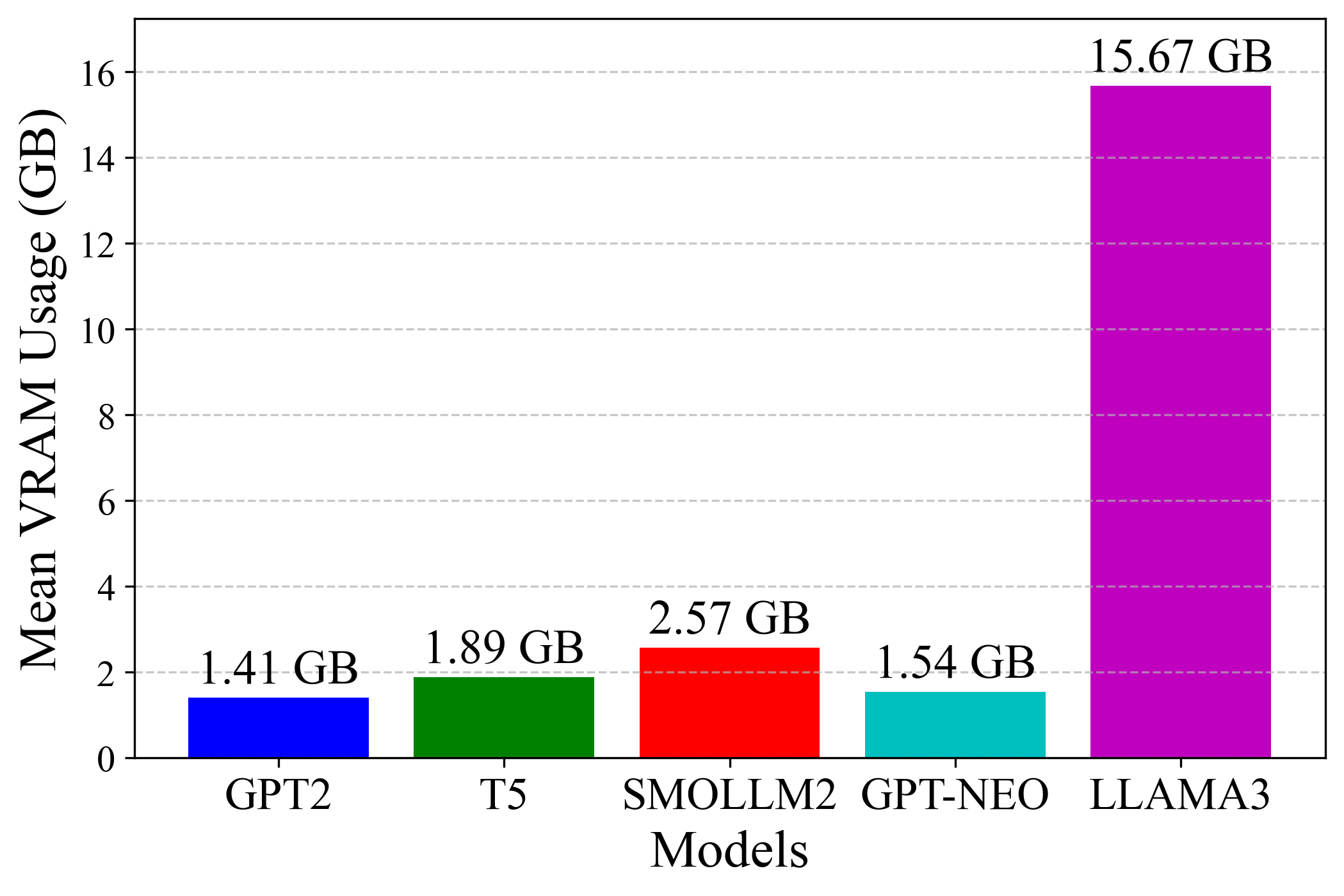}\\
    \small (b) Average VRAM usage
\end{minipage}
\caption{Comparison of (a) trainable parameter percentages and (b) average VRAM usage in training.}
\label{fig:resources}
\vspace{-20pt} 
\end{figure}

The primary outcome of the proposed \gls{lora} approach is the reduction of training parameters in finetuning the \glspl{slm} to achieve improved \gls{bbr} performance. 
Fig. \ref{fig:resources}(a) demonstrates a comparison of training parameters in \glspl{slm} of interest and LLaMA 3.2, a \gls{llm} with more than 3~B parameters, which we have used to benchmark the performance of the fine-tuned \glspl{slm} for the rest of this work. 
Respectively, only $9.44\%$ GPT-2's 137~M parameters, $5.97\%$ of 237~M parameters in T5, $5.35\%$ out of 132~M parameters in GPT-Neo, and $6.76\%$ of SmolLM2's 388~M parameters are leveraged in training with the \gls{lora} method. 
Comparatively, $1.13\%$ out of 3~B parameters in LLaMA 3.2 is fine-tuned with the proposed method, thus reflecting on the significant reduction of the adapted trainable parameters. 
In addition, \gls{vram} utilization in training is significantly reduced through the \gls{lora} approach as depicted in Fig. \ref{fig:resources}(b). 
As expected, \glspl{slm}' \gls{vram} usage is significantly lower compared to the LLaMA 3.2 model, which recoded 15.67~GB of mean \gls{vram} consumption. 
Moreover, all four \glspl{slm} recorded a \gls{vram} usage below 2.57~GB, SmolLM2 registering the highest and GPT-2 with the lowest of 1.41~GB.



\begin{figure*}[h]
\centerline{\includegraphics[width=2\columnwidth]{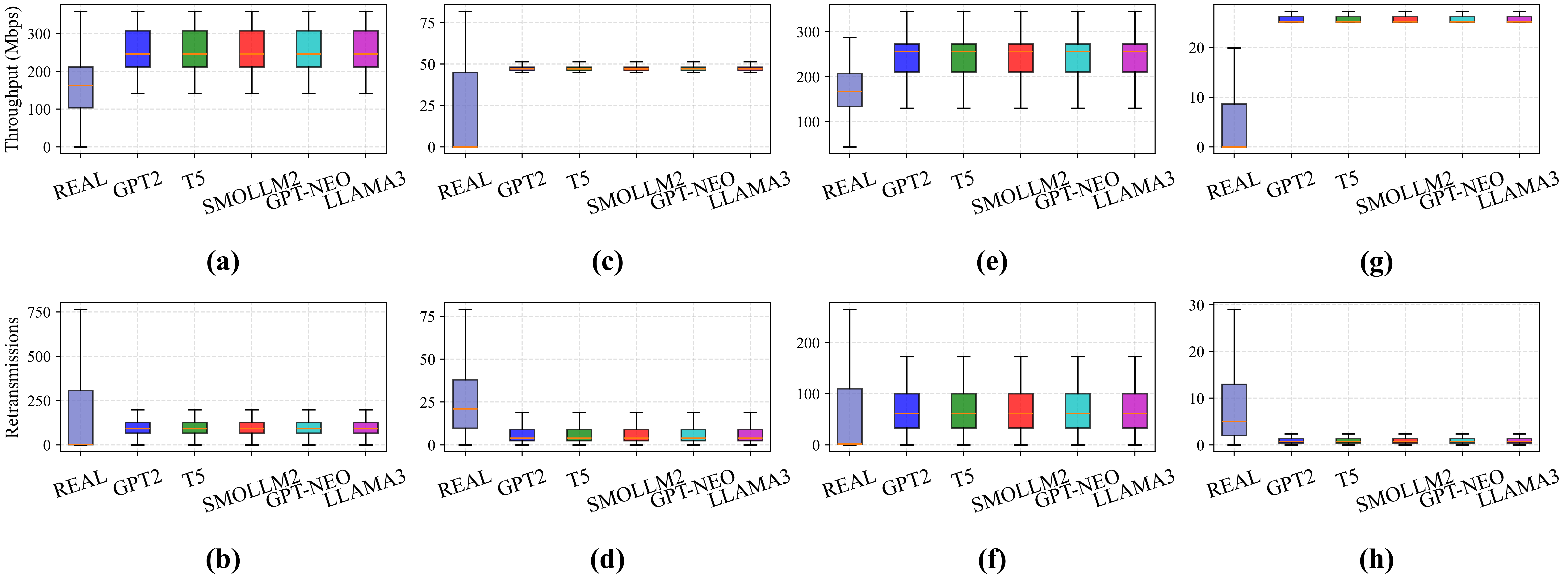}}
\caption{
Evaluation of SLM-predicted BBR pacing gains in terms of throughput and retransmissions, compared against native Tokyo BBR flows and the LLaMA~3.2 predictions.
(a) Downlink individual flow throughput.
(b) Downlink individual flow retransmissions.
(c) Uplink individual flow throughput.
(d) Uplink individual flow retransmissions.
(e) Downlink competing flow throughput.
(f) Downlink competing flow retransmissions.
(g) Uplink competing flow throughput.
(h) Uplink competing flow retransmissions.
\label{fig:results}}
\vspace{-15pt} 
\end{figure*}

\begin{figure}[h]
\centerline{\includegraphics[width=0.6\columnwidth]{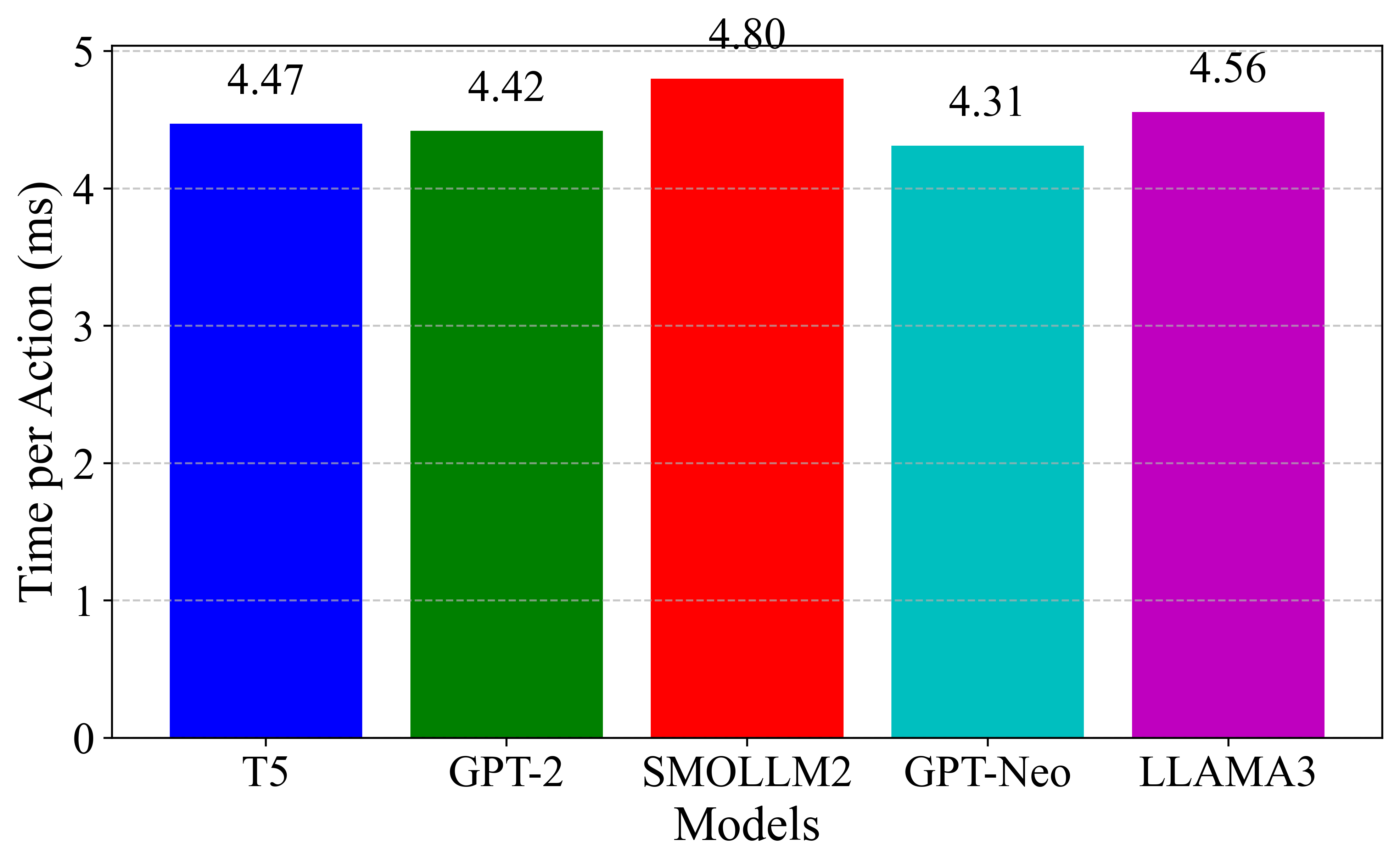}}
\caption{Action generation time comparison in inference.  \label{fig:action_time}}
\vspace{-14pt} 
\end{figure}

\vspace{-2mm}
\subsection{Throughput Model}

To evaluate the impact of fine-tuned \glspl{slm} on pacing gain decisions in \gls{bbr} congestion control flows, we construct a surrogate approach that estimates the resulting throughput and retransmissions without requiring full \gls{tcp}-in-the-loop replays. 
The surrogate model relies on the measured \texttt{iperf3} statistics from the \gls{bbr} flows, and \gls{slm}/\gls{llm}-generated discrete pacing-gain actions. 
This approach enables scalable, model-agnostic evaluation of control policies.

\begin{equation}
    b_{\text{send}}(i) = G(a_i)\,\widehat{B}_\text{w}(i),
\end{equation}
A simplified version of the \gls{bbr} sending rate can be given as the above equation, where $\widehat{B}_\text{w}(i)$ is the estimated bottleneck bandwidth, approximated by a 95-th percentile rolling window value $B_{\text{cap}}(i)$.  
Thus, the delivered throughput through \gls{slm} predicted pacing gains can be given as:
\begin{equation}
    T^{\text{SLM}}(i)
    = 
    B_{\text{cap}}(i)
    \cdot 
    \min\!\bigl(G(a_i),~1.0\bigr).
    \label{eq:throughput-surrogate}
\end{equation}

In Eq. (\ref{eq:throughput-surrogate}), since probe-up or cruise resulted in $G(a_i) \ge 1$, throughput saturates at the bottleneck, and throughput is proportionally reduced in probe-down ($G(a_i) < 1$).
When $G(a_i) > 1$, the sender transmits at a rate higher than the bottleneck capacity, even though this increases the amount of data in flight, the delivered throughput remains constrained by the bottleneck bandwidth. 
Consequently, the excess traffic accumulates in the bottleneck queue, leading to increased queuing delay and a higher probability of packet loss and retransmissions. 
Therefore, gains greater than one primarily manifest as elevated \gls{rtt} and retransmissions rather than proportional throughput improvements. 

\vspace{-4mm}
\subsection{Retransmission Model}

To evaluate the impact of the fine-tuned language model predicted pacing gains on retransmission, we introduce a phase-aware methodology that maps the predicted pacing gain to an expected retransmission level. 
To model this effect smoothly, we define a shifted softplus activation approach $(\operatorname{softplus}(x) = \ln(1 + e^x))$. 
The probe strength can be computed as:
\begin{equation}
\label{eq:probe_strength}
S(G_i) = \max\!\left( 
\operatorname{softplus}(\beta(G_i - 1)) - \operatorname{softplus}(0),\; 0 
\right),
\end{equation}
which guarantees $S(G_i) = 0$ when $G_i = 1$, ensuring no artificial penalty in cruise mode.
The normalized probe-up component is defined as:
\begin{equation}
\label{eq:probe_term}
\Phi(G_i) = 
\left(
\frac{S(G_i)}{S(G_{\max})}
\right)^{\alpha}
\end{equation}
where $G_{\max}$ is the maximum allowed pacing gain, parameters $\alpha = 1.5$ controls superlinear growth, and $\beta =5$ controls sharpness.
In the absence of probing, non-zero retransmissions occur due to wireless impairments and background contentions. 
This is represented through a small baseline term $\epsilon, (\epsilon \ll 1)$, where it sets the minimum loss floor.

Therefore, the instantaneous loss factor can be defined as:
\begin{equation}
\label{eq:loss_factor}
L_i = U_i \left[ \epsilon + (1-\epsilon)\Phi(G_i) \right].
\end{equation}
ensuring retransmissions increase only when both utilization and probe aggressiveness are high.

For pacing gains below unity ($G_i < 1$), \gls{bbr} intentionally drains queues. This effect is explicitly rewarded by reducing the loss factor:
\begin{equation}
\label{eq:probe_down}
L_i \leftarrow L_i \left( 1 - \kappa_{\text{down}}(1 - G_i) \right),
\end{equation}
where $\kappa_{\text{down}} \in (0,1)$ controls the strength of the probe-down benefit.
However, the final loss factor is limited to a valid range  as $L_i = \min\left( \max(L_i, 0), 1 \right)$
Therefore, the estimated retransmissions are obtained by scaling within the local congestion envelope:
\begin{equation}
\label{eq:retrans}
\tau_i^{\text{SLM}} = \tau_{\min} + \bigl(\tau_{\max,i} - \tau_{\min}\bigr)\, L_i
\end{equation}
where $\tau_{\max,i}$ is the rolling maximum of retransmissions, capturing the local congestion regime. 




As discussed in \S~\ref{sec:testbed}, we captured \gls{bbr} \texttt{iperf3} data for independent downlink, uplink streams, and parallel competitive downlink and uplink streams, which compete against another three \glspl{cca}: Hybla, Cubic, and Vegas across six cities. 
We leveraged Ohio, São Paulo, London, Mumbai, and Sydney \gls{bbr} data to finetune the \glspl{slm} and LLaMA 3.2 \gls{llm}, and evaluate them with the Tokyo data set. 
\gls{slm} predicted \gls{bbr} pacing gain driven results for Tokyo are then compared against the actual data and LLaMA 3.2 model. 
Calculated throughput and retransmissions through the surrogate models discussed above are depicted in Fig. \ref{fig:results}. 
Throughput and retransmission comparisons of the downlink and uplink for independent \gls{bbr} streams are given in Fig.~\ref{fig:results}(a), Fig.~\ref{fig:results}(b), Fig.~\ref{fig:results}(c), and Fig.~\ref{fig:results}(d), respectively. 
Fig.~\ref{fig:results}(e), Fig.~\ref{fig:results}(f), Fig.~\ref{fig:results}(g), and Fig.~\ref{fig:results}(h) present the throughput and retransmission comparisons of competing downlink and uplink \gls{bbr} streams. As shown in Fig.~\ref{fig:action_time}, we observe that the proposed \gls{lora} approach reduces language model inference latency below 5~ms, which requires further extensive evaluation.


\vspace{-2mm}
\section{Conclusion}

We present a framework to adapt \gls{slm} to improve \gls{bbr} \gls{cca} for \gls{leo} satellite Internet. 
Through a globally distributed testbed over SpaceX’s Starlink, we present an empirical evaluation of \gls{bbr} characteristics in \gls{leo} satellite networks, highlighting its throughput advantage over other \glspl{cca} despite elevated retransmission rates.
As a solution to the highlighted drawback, this work presents a \gls{slm} driven smooth pacing gain approach, reducing the inherent aggressiveness in \gls{bbr}. 
Using four \glspl{slm}, we develop reward-maximizing expert action, evaluate the proposed approach and benchmark the results with a fine-tuned \gls{llm} and real \gls{bbr} flows. 
Results evaluated via surrogate models reveal that the distilled \glspl{slm} retain throughput while reducing the retransmissions through intelligent pacing gain selection. 
Due to the lightweight deployment advantages offered by \glspl{slm}, our developed model represents a further step toward intelligence-driven network control over satellite Internet.
\\
\\


\vspace{-30pt} 
\bibliographystyle{IEEEtran}
\bibliography{reference}

\vfill

\end{document}